 \documentclass[final,3p,times,twocolumn]{elsarticle}

\usepackage{graphicx}

\usepackage{amssymb}

\usepackage{url}

\journal{Nuclear Instruments and Methods in Physics Research A}

\begin{document}

\begin{frontmatter}



\title{Characterization of large area APDs for the EXO-200 detector}


\author[stanford]{R.~Neilson\corref{cor}}
\cortext[cor]{Corresponding author. Tel: +1-650-725-6342; fax: +1-650-725-6544.}
\ead{rneilson@stanford.edu} 
\author[stanford]{F.~LePort}
\author[stanford]{A.~Pocar\fnref{andrea}}
\author[umass]{K.~Kumar}
\author[slac]{A.~Odian}
\author[slac]{C.Y.~Prescott}
\author[stanford]{V.~Tenev}
\author[slac]{N.~Ackerman}
\author[itep]{D.~Akimov}
\author[bern]{M.~Auger}
\author[csu]{C.~Benitez-Medina}
\author[slac]{M.~Breidenbach}
\author[itep]{A.~Burenkov}
\author[slac]{R.~Conley}
\author[csu]{S.~Cook}
\author[stanford]{R.~deVoe}
\author[stanford]{M.J.~Dolinski}
\author[csu]{W.~Fairbank Jr.}
\author[laurentian]{J.~Farine}
\author[stanford]{P.~Fierlinger\fnref{peter}}
\author[stanford]{B.~Flatt}
\author[bern]{R.~Gornea}
\author[stanford]{G.~Gratta}
\author[stanford]{M.~Green}
\author[maryland]{C.~Hall}
\author[csu]{K.~Hall}
\author[laurentian]{D.~Hallman}
\author[carleton]{C.~Hargrove}
\author[slac]{S.~Herrin}
\author[slac]{J.~Hodgson}
\author[maryland]{L.J.~Kaufman}
\author[itep]{A.~Kovalenko}
\author[bama]{D.S.~Leonard\fnref{doug}}
\author[slac]{D.~Mackay}
\author[csu]{B.~Mong}
\author[stanford]{M.~Montero Diez}
\author[bama]{E.~Niner}
\author[stanford]{K.~O'Sullivan}
\author[bama]{A.~Piepke}
\author[slac]{P.C.~Rowson}
\author[carleton]{D.~Sinclair}
\author[slac]{K.~Skarpaas}
\author[maryland]{S.~Slutsky}
\author[itep]{V.~Stekhanov}
\author[carleton]{V.~Strickland}
\author[stanford]{K.~Twelker}
\author[bern]{J.-L.~Vuilleumier}
\author[slac]{K.~Wamba}
\author[laurentian]{U.~Wichoski}
\author[slac]{J.~Wodin}
\author[slac]{L.~Yang}
\author[maryland]{Y.-R.~Yen}
\fntext[andrea]{Now at University of Massachsetts, Amherst MA, USA.}
\fntext[peter]{Now at Technical University Munich, Munich, Germany.}
\fntext[doug]{Now at University of Maryland, College Park MD, USA.}

\address[stanford]{Physics Department, Stanford University, Stanford CA, USA}
\address[umass]{Physics Department, University of Massachsetts, Amherst MA, USA}
\address[slac]{SLAC National Accelerator Laboratory, Stanford CA, USA}
\address[itep]{Institute for Theoretical and Experimental Physics, Moscow, Russia}
\address[bern]{LHEP, Physikalisches Institut, Univerisity of Bern, Bern, Switzerland}
\address[csu]{Physics Department, Colorado State University, Fort Collins CO, USA}
\address[laurentian]{Physics Department, Laurentian University, Sudbury ON, Canada}
\address[maryland]{Physics Department, University of Maryland, College Park MD, USA}
\address[carleton]{Physics Department, Carleton Univerisity, Ottawa ON, Canada}
\address[bama]{Department of Physics and Astronomy, University of Alabama, Tuscaloosa AL, USA}

\begin{abstract}

EXO-200 uses 468 large area avalanche photodiodes (LAAPDs) for detection of scintillation light in an ultra-low-background liquid xenon (LXe) detector. We describe initial measurements of dark noise, gain and response to xenon scintillation light of LAAPDs at temperatures from room temperature to 169~K---the temperature of liquid xenon. We also describe the individual characterization of more than 800 LAAPDs for selective installation in the EXO-200 detector.

\end{abstract}


\end{frontmatter}



\section{Introduction}

In recent years Large Area Avalanche PhotoDiodes (LAAPDs) have been discussed as photodetectors of scintillation light \cite{lopes01,solovov02,moszynski03b}. LAAPDs are compact, semiconductor devices with high quantum efficiency (QE) from the infrared to the vacuum ultraviolet (VUV). Made mostly of high purity silicon, very lightweight, and fabricated in a clean room environment, they can be produced with low intrinsic radioactive contamination \cite{leonard08}. Hence they are well suited to the detection of scintillation light in ultra-low background experiments, especially when the wavelength to be detected is in regions of the spectrum that are problematic for photomultiplier tubes (PMTs). This is the case for noble elements as the scintillation light is in the VUV region \cite{jortner65}. Although several groups working on the detection of low energy, rare events have been investigating LAAPDs \cite{ni05,chandrasekharan06}, EXO-200 is the first such detector to make wide use of them.

\section{The use of LAAPDs in EXO-200}

EXO (Enriched Xenon Observatory) is a program aimed at building a ton-class neutrinoless double beta ($0\nu\beta\beta$) decay \cite{avignone08} detector using xenon enriched to 80\% in the isotope 136 as the source and detection medium \cite{danilov00}. While the EXO collaboration is planning to build a ton-scale detector with the ability to retrieve and identify the $^{136}$Ba atom produced in the double beta decay of $^{136}$Xe, an intermediate scale (200~kg of enriched xenon, 80\% $^{136}$Xe) detector, without the barium tagging feature, EXO-200, is currently close to the data taking phase. 

The xenon in EXO-200 is kept in liquid phase at a temperature $\sim$170~K and a pressure $\sim$1.5~atm. The two electrons produced in the double beta decay are detected in an electric field of 1-4 kV/cm by the simultaneous readout of ionization and scintillation. This technique has been shown to provide better energy resolution than can be achieved with ionization or scintillation detection alone \cite{conti03,aprile07}, as required to suppress backgrounds by discriminating the sharp peak of the $0\nu\beta\beta$ decay from other events with broader energy spectra. In the EXO-200 Time Projection Chamber (TPC), the ionization signal is collected by crossed-wire
grids, and the $\sim$178~nm wavelength  VUV scintillation light is detected by LAAPDs, hence measuring the total energy of the decay and its three-dimensional location. The third dimension is provided by the drift time, using the scintillation signal as the start time. The liquid xenon (LXe) detector is contained in a low-activity thin-walled (1.5~mm thick) copper vessel in the shape of a cylinder about 40~cm long and 40~cm in diameter. The copper vessel is submerged in HFE-7000\footnote{Novec Engineered Fluid HFE-7000 (C$_3$F$_7$OCH$_3$) is a hydrofluoroether heat transfer fluid made by 3M, St. Paul MN, USA.} contained in a low-activity copper cryostat. The HFE-7000 is an ultra-clean fluid that is in liquid phase in a broad range of temperatures, encompassing 300 and 170~K. This fluid is used as the innermost $\gamma$ and neutron shield, and as the thermal bath to maintain a uniform temperature around the LXe vessel.

\begin{figure}
\includegraphics[width=78mm]{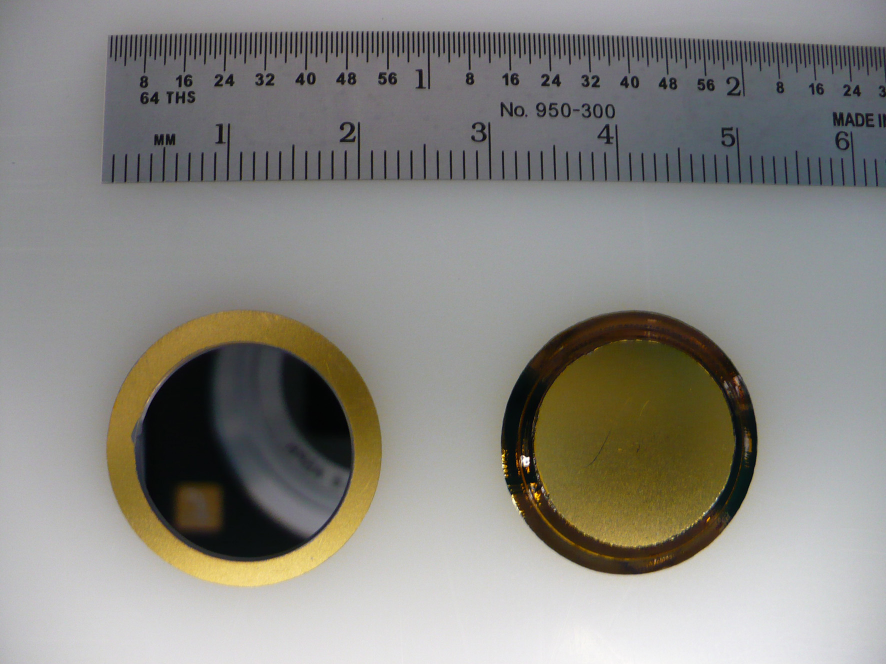}
\caption{API LAAPD. The ruler indicates the size in cm and inches. The gold-plated cathode is displayed on the right and the active surface surrounded by the gold-plated ring-wafer anode on the left. Note the bevel on the cathode side, providing a longer distance along which to hold the bias high voltage.}
\label{fig:apd_picture}
\end{figure}

\begin{figure}
\includegraphics[width=78mm]{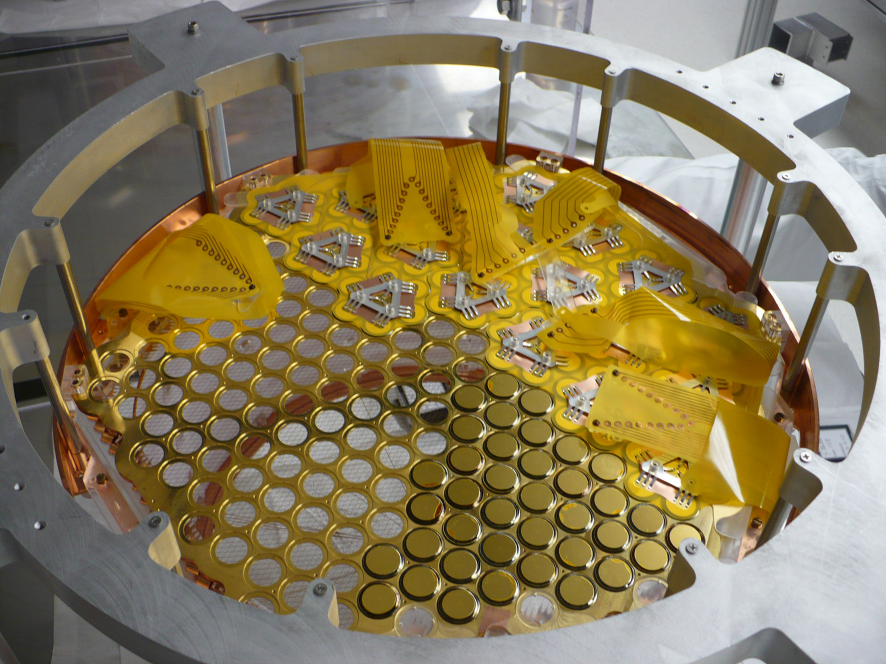}
\caption{A partially filled LAAPD platter. The triangular platinum-plated contact springs have been installed at the top and right; the LAAPDs at the bottom center have yet not had their contact springs installed.}
\label{fig:platter}
\end{figure}

The EXO-200 LAAPDs are used ``bare'', i.e. without standard encapsulation, and immersed in LXe, which acts as an excellent electrical insulator. This eliminates issues of radioactive background and differential expansion related to the use of the ceramic encapsulation. In addition it allows for a higher packing density and eliminates the front window, resulting in a higher efficiency for VUV photon detection. The LAAPDs are Advanced Photonix (API) unencapsulated dies\footnote{API product \# SD630-70-75-500 (\url{http://www.advancedphotonix.com/ap_products/pdfs/SD630-70-75-500.pdf}) by Advanced Photonix Inc., 1240 Avenida Acaso, Camarillo, CA 93012, USA.}, shown in Figure~\ref{fig:apd_picture}. They have 16~mm diameter active area (200~mm$^2$), with an overall physical diameter between 19.6 and 21.1~mm.  The thickness varies between 1.32 and 1.35~mm and the mass is 0.5~g. The LAAPDs are mounted on two gold-plated copper platters (Figure~\ref{fig:platter}), one at each end of the TPC. The 234 devices on each platter are arranged in a triangular grid with a photosensitive packing fraction of 48\%. There are no LAAPDs on the sides of the detector because of the large electric field;  Teflon\footnote{Dupont TE-6472, a variety of {\it modified} polytetrafluoroethylene (PTFE).} reflectors are mounted in these regions. Teflon is known to be a good diffuse reflector of xenon scintillation light \cite{yamashita04}, increasing the light collection efficiency by 50--150\%, depending on the location of the event in the detector. The front face of the LAAPD support platter is beveled and coated\footnote{Coating done by Vacuum Process Engineering, Inc., 110 Commerce Circle, Sacramento, CA. 95815.} with aluminum and magnesium fluoride in order to reflect photons that do not hit an LAAPD, further increasing light collection efficiency. The LAAPDs are electrically ganged in groups of seven (or, in some special cases, five or six), giving a total of 74 LAAPD electronic readout channels (37 per side). Each group is held in place by a platinum-plated phosphor bronze spring, which also provides electrical contact to the gold-plated cathodes of the LAAPDs.

LAAPDs are silicon devices consisting of a p-n junction. Light photons are absorbed on the active surface in the p-region of the diode and produce electron-hole pairs. The electric field drives the primary electrons to a multiplication region, where the field strength is above the threshold for the production of new electron-hole pairs by impact ionization. Multiplication gains of order 100 are typical, increasing more than exponentially with the applied voltage. The devices used in EXO-200 are made by growing the p-type epitaxial layer on n-type neutron transmutation doped silicon~\cite{moszynski02}. A second silicon wafer, ring-shaped and gold-plated, is bonded by a layer of aluminum to form the anode. The external contact of the cathode ring is also gold plated. The edge of the device is beveled and coated with a polyimide paint to improve the breakdown characteristics at the edges. Extensive radioactivity measurements were made of both complete and partial LAAPDs~\cite{leonard08}. Following these measurements, the aluminum used to bond the ring wafer to the rest of the device was found to be unsatisfactory and was replaced with a special type of ultra-pure aluminum\footnote{Hydro Aluminium Deutchland GmbH, Bonn, Germany.}.

\section{LAAPD handling and initial qualification}

A total of 851 production LAAPDs were purchased for EXO-200 and produced by API between September 2006 and June 2008. Great care has been taken at all times in the handling of these LAAPDs, both to protect the unencapsulated diodes from damage and to maintain the ultra-low radioactive background and xenon purity requirements of EXO-200. In particular, unencapsulated diodes are adversely affected by humidity, so the devices are transported in a dry-nitrogen filled container and stored in a dry box purged with boil-off nitrogen. All testing of the production LAAPDs is done in a clean room.

A number of laboratory tests were performed to characterize the properties of unencapsulated LAAPDs. In particular, effort was put into understanding the noise, gain and QE of the devices, and their behavior as a function of temperature. This initial testing of pre-production LAAPDs, not to be installed in EXO-200, was done in a single-device setup with a commercial integrating preamplifier, shaper and multichannel analyzer\footnote{ORTEC 142H preamplifier, ORTEC 672 amplifier and PerkinElmer Trump-PCI-2K PC-based multichannel analyzer.}. In addition, all 851 production LAAPDs have been individually characterized under expected operating conditions. This was done in a multiple-LAAPD test setup with custom electronics.

\section{Multiple-LAAPD test setup}

\begin{figure}
\includegraphics[width=78mm]{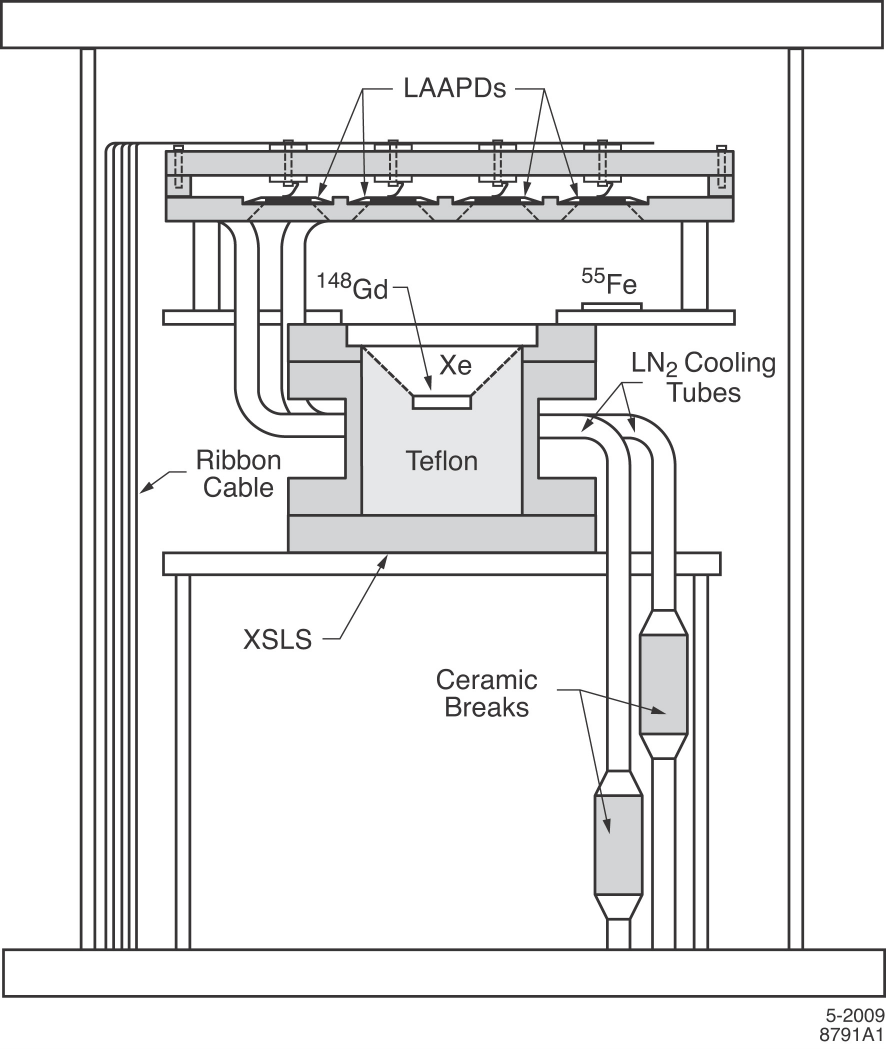}
\caption{Test chamber design. VUV photons from the xenon scintillation light source (XSLS) and X-rays from the $^{55}$Fe source are detected by the downward facing LAAPDs.}
\label{fig:test_chamber}
\end{figure}

\begin{figure}
\includegraphics[width=78mm]{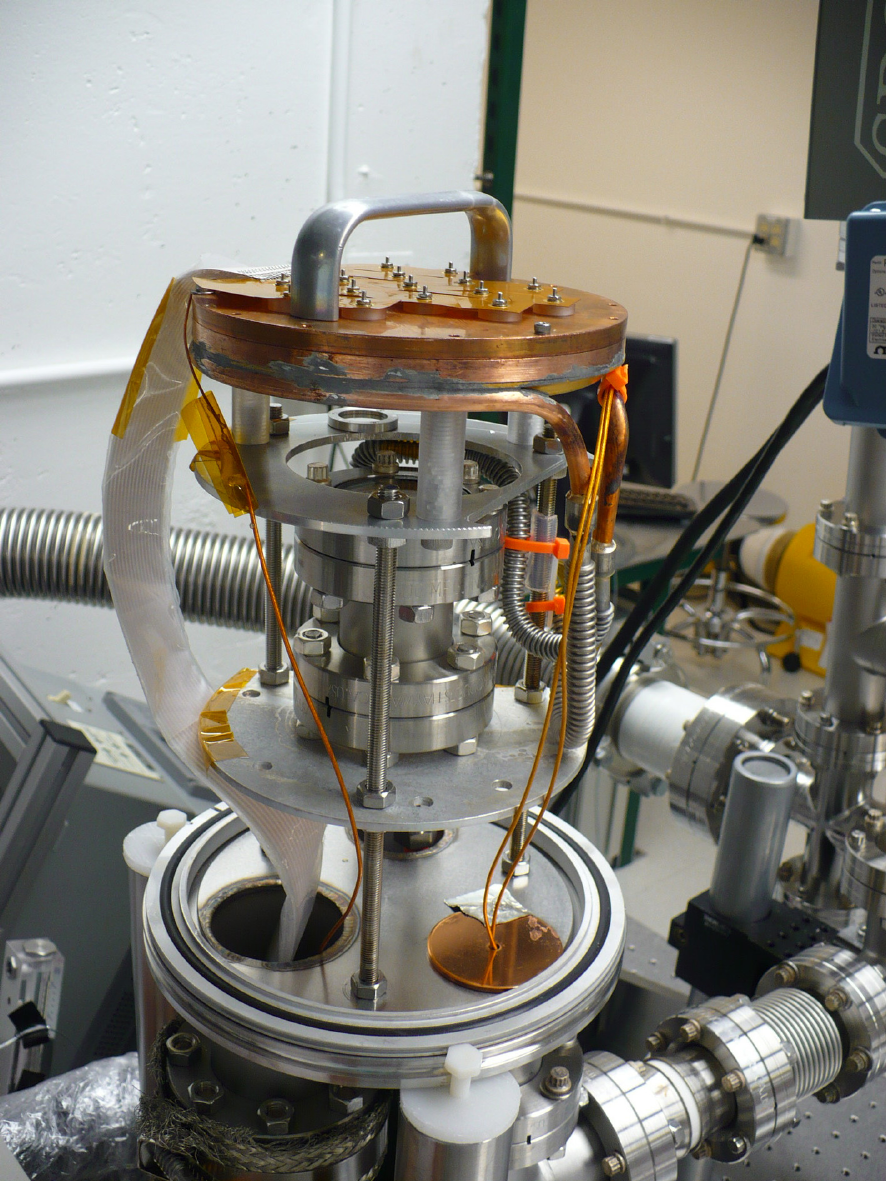}
\caption{The vacuum chamber of the multiple-LAAPD test setup. The copper mounting disks are at the top of the picture. The XSLS is in the middle, the cooling tubes on the right, and the 16 conductor flex cable on the left. The pumping manifold can be seen at the bottom right and in the background.}
\label{fig:test_system}
\end{figure}

\begin{figure}
\includegraphics[width=78mm]{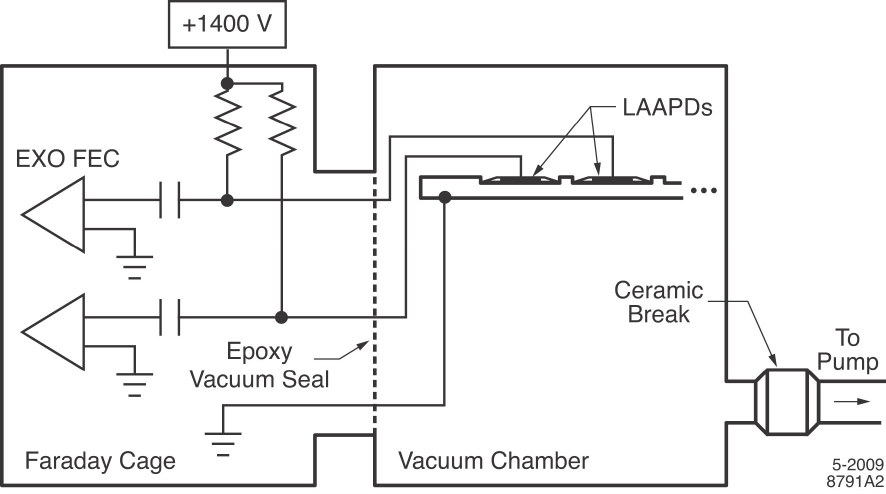}
\caption{Electrical schematic of the test system up to the preamplifiers on the FEC.}
\label{fig:electronics}
\end{figure}

The testing and characterization of the production LAAPDs is done in a vacuum chamber maintained at better than $10^{-6}$~torr by a turbomolecular pump backed by a dry scroll pump. Figure~\ref{fig:test_chamber} shows a schematic of the test chamber and Figure~\ref{fig:test_system} displays a photograph of the system. Sixteen devices, mounted face down on a 9.5~mm thick copper disk, are tested simultaneously in the chamber. The copper disk connects together the anodes for all 16 LAAPDs and maintains them at a constant and equal temperature. The back surface of each device is contacted by a beryllium copper spring, mounted, with nylon supports, on a second, removable copper disk. These cathode connections reverse bias each device at $\sim$1400~V for a gain of $\sim$100. A 16 conductor flex cable transmits signals to the front-end electronics card (FEC) outside the vacuum chamber (Figure~\ref{fig:electronics}).

The temperature of the devices is maintained near the operating temperature of EXO-200 (170~K) for the testing.  The LAAPDs are cooled by boil-off nitrogen gas, flowing through a copper tube brazed to the circumference of the lower copper mounting disk. The temperature of the nitrogen gas is regulated by a two-stage resistive heater attached to a section of the cooling tube external to the vacuum chamber. A 150~W heater provides coarse temperature control. A 20~W heater, controlled by a PID feedback circuit reading out a resistance temperature detector (RTD) on the LAAPD mounting disk, provides fine temperature control. This configuration achieves cool down from room temperature to 170~K in approximately two hours and temperature stability better than 0.2~K. It is important that the temperature be held within a very tight range because of the strong dependence of LAAPD gain on temperature, and a fast cool-down is important because of the production nature of the measurements. Ceramic insulators are used to electrically isolate the vacuum chamber from the boil-off nitrogen source and from the vacuum pump system.

Two radioactive sources mounted below the copper disk are used to measure certain parameters of the LAAPDs. A 5~$\mu$Ci $^{55}$Fe source provides 5.9~keV X-rays to measure the absolute gain of the devices, by total absorption directly in the junction. X-rays incident on the device junction generate electron-hole pairs, with an average of one pair produced for each 3.66~eV of incident energy \cite{scholze00}. Thus, a 5.9~keV X-ray from the $^{55}$Mn de-excitation generates $\sim$1600 electron-hole pairs in the active region. Measuring the gain in this way assumes that the gain is linear in the number of electron-hole pairs produced. This is not completely correct, as X-rays produce high charge densities, decreasing the local electric field and producing local heating. The non-linearity of the gain has been measured \cite{moszynski02}, and the difference in gain between 5.9~keV X-rays and VUV photons is less than 1\% at a gain of 100.

A xenon scintillation light source (XSLS), made by enclosing a 1~nCi $^{148}$Gd alpha source within a cell filled to 3.3~atm with xenon gas, produces VUV photons. The scintillation light resulting from the alpha decays produces the characteristic spectrum of xenon, peaked at 178~nm (7~eV photons), so that the QE can be measured in the most relevant way. The top of the cell is fitted with a quartz window and the scintillation light generated is enhanced by a conical Teflon reflector inside the xenon cell. While $^{55}$Fe events deposit their energy into individual LAAPDs, events from the XSLS consist of a broad fan of VUV photons producing signals in all LAAPDs. Therefore the QE measure is a relative one, normalized to a ``standard'' LAAPD that is installed in the central position.

\subsection{Data acquisition}

The multiple-LAAPD test setup is read out using a partial prototype of the EXO-200 Data Acquisition (DAQ) system. Signals from the LAAPDs are digitized on a custom-made 16 channel Front End Card (FEC). Each channel consists of a pre-amplifier, an analog shaping circuit, and a 12 bit charge analog-to-digital converter (ADC). The output of the preamplifiers is sampled and digitized at 1~MHz. The digitized output from all 16 channels is transmitted over an optical cable to a Trigger Electronics Module (TEM). This module maintains a buffer of 1024 digitization samples and stores the trigger logic. Once a trigger is issued an event is created, consisting of 2048 samples from each of the 16 channels, centered on the time of the trigger. Thus each event contains approximately 2~ms of data. The event data is stored on a computer and is analyzed offline. The data acquisition rate is limited to 80 events per second by the transfer rate to the computer.

Since the two radioactive sources produce different signatures, the DAQ is triggered separately for $^{55}$Fe events and XSLS events. An $^{55}$Fe trigger is issued if the signal from any one channel crosses a pre-defined threshold and an XSLS trigger is issued if the sum of all 16 channels crosses a threshold.

\subsection{Test procedure}

All production LAAPDs are characterized for leakage current, noise, gain and QE in sets of 15. In each batch, the central (16th) location on the copper support is occupied by an LAAPD designated as ``standard'' and left in place for many measurement cycles.   This LAAPD provides information on the system stability.

The LAAPDs are first subjected to a break-in cycle consisting of at least two cool-downs from room temperature to 169~K, one warm-up from 169~K to room temperature, and 24 hours maintained at 169~K. During this break-in, the leakage currents of the devices are measured, both at room temperature and at 169~K. Following this, measurements at 169~K of noise, gain and QE require the accumulation of a few thousand events per device. The full data acquisition for a set of 16 LAAPDs takes 2--3 hours once the temperature has stabilized.

Of the 851 LAAPDs, about 180 were found to be either unstable or very noisy at 169~K and biased for a gain of 100. It was not possible to conduct the full suite of measurements on these devices and they are not included in the noise, gain and QE results presented in this paper.

\section{Noise}

\begin{figure}
\includegraphics[angle=90,width=78mm]{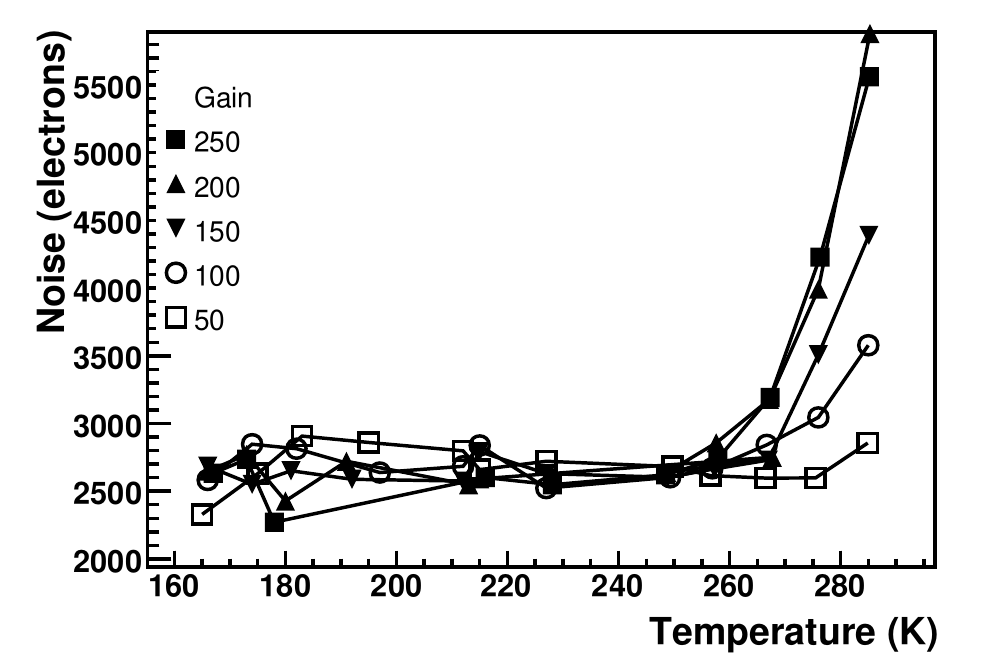}
\caption{Dark noise as a function of temperature for fixed gains from 50 to 250 for one specific LAAPD. The base value of $\sim$2600 electrons is the result of the capacitance on the input of the preamplifier; the excess noise above 250 K is due to leakage current.}
\label{fig:noise_vs_temp}
\end{figure}

Figure~\ref{fig:noise_vs_temp} shows the measured dark noise as a function of temperature for fixed gains of 50, 100, 150, 200 and 250 for a typical LAAPD in the single-device setup. Electronic noise has been subtracted, and the RMS widths are given in electron equivalent charge. The noise performance of the device improves with decreasing temperature down to $\sim$250~K, with no observable change below that.

The intrinsic device noise includes contributions from both the capacitance and the leakage current. We have investigated each of these separately. The capacitance, responsible for part of the noise, decreases markedly as the bias voltage increases because of the change in depletion depth.    Typical devices have a capacitance of $\sim$1~nF at 50~V and $\sim$200~pF at 1400~V.   

\begin{figure}
\includegraphics[angle=90,width=78mm]{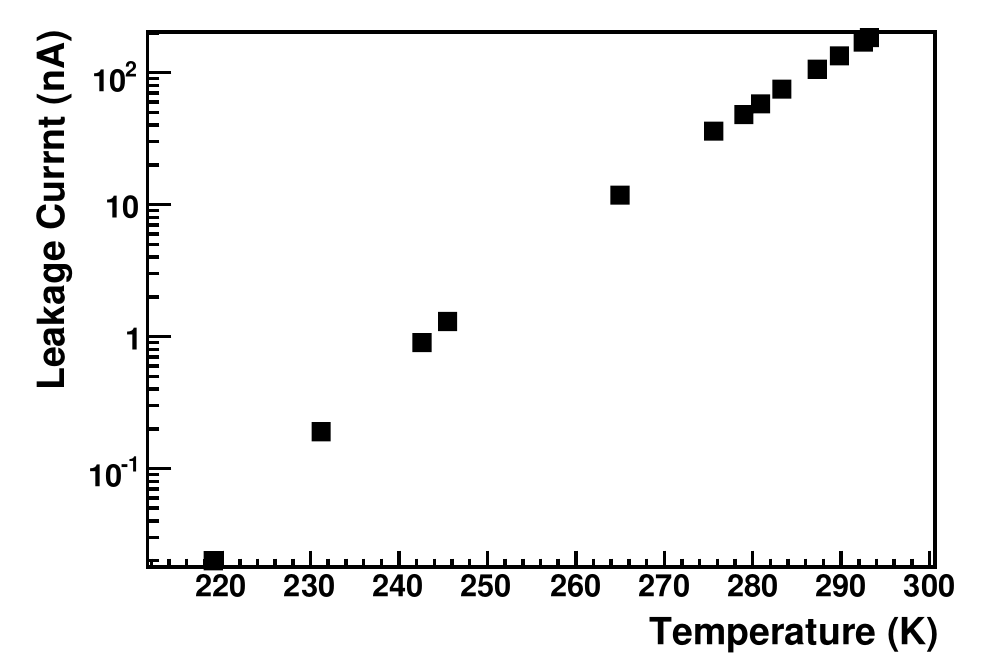}
\caption{Leakage current as a function of the temperature for a typical device at fixed gain of 100.}
\label{fig:leakage_current}
\end{figure}

Figure~\ref{fig:leakage_current} shows the leakage current as a function of temperature for a single LAAPD. The leakage current drops from $\sim$300~nA at room temperature to $\sim$20~pA at 218~K, a drop of over four orders of magnitude. This is the primary reason the device noise improves with decreasing temperature, with leakage current typically becoming a negligible contribution to the noise below 250~K. However, we have observed many devices with anomalously high leakage current.   All devices were tested by the manufacturer at room temperature to ensure that the leakage current at a gain of 100 does not exceed 1~$\mu$A.  Despite this, our testing found a substantial number of devices failing this criterion, indicating deterioration after the testing at the manufacturer's facility.    While it is unclear what causes such deterioration, we have observed no significant change in the leakage current after the LAAPDs are stored in our facility. As part of the acceptance for EXO-200 we have also measured leakage currents at 169~K, a more interesting parameter that the manufacturer was unable to test.   A number of devices exhibit leakage currents much greater than 1~nA at 169~K, in some cases with no indication of high leakage current at room temperature. We have found that anomalously high leakage current at 169~K is often correlated with noisy or unstable devices.

\begin{figure}
\includegraphics[angle=90,width=78mm]{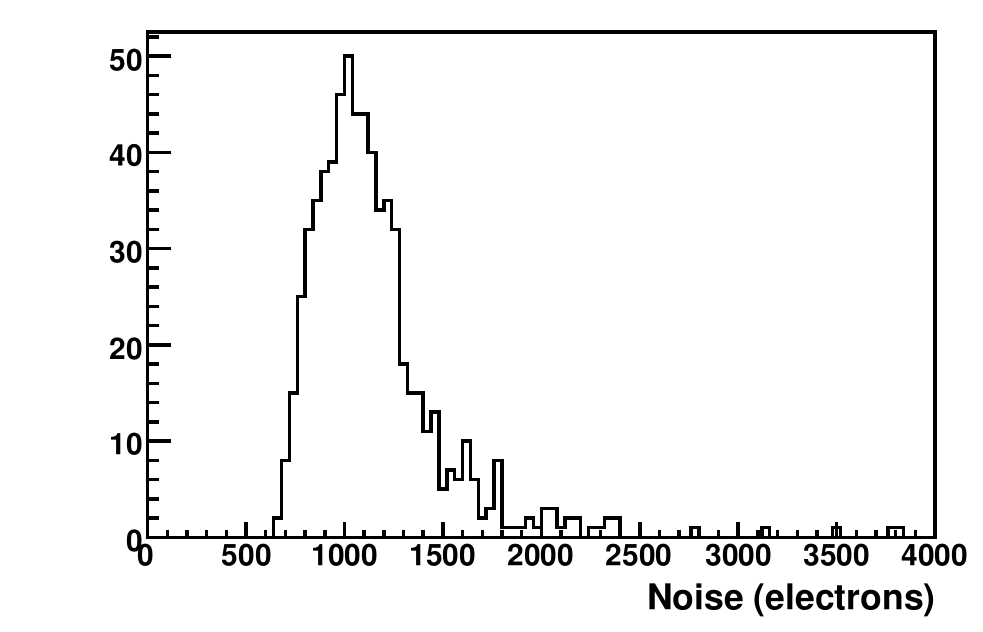}
\caption{Noise of the EXO-200 production LAAPDs measured at 169~K and a gain of 100. A total of 667 devices are represented in the histogram.}
\label{fig:apd_noise}
\end{figure}

In the 16 device setup, the noise of the LAAPDs is determined by an analysis of the ADC output when there is no trigger. The RMS of the output around the baseline is calculated for a large number of events and averaged to give a dark noise measurement, shown in Figure~\ref{fig:apd_noise} for all devices. The noise of most devices lies between 700 and 1500 electrons.   Although in this case the contribution of the electronics has not been subtracted, the measured noise is substantially lower than that measured in the single-device setup because the preamplifier is better matched to the capacitance of these LAAPDs.   Devices showing electronic noise greater than 1500 electrons are considered unsuitable for installation in EXO-200.

\section{Energy resolution}

\begin{figure}
\includegraphics[angle=90,width=78mm]{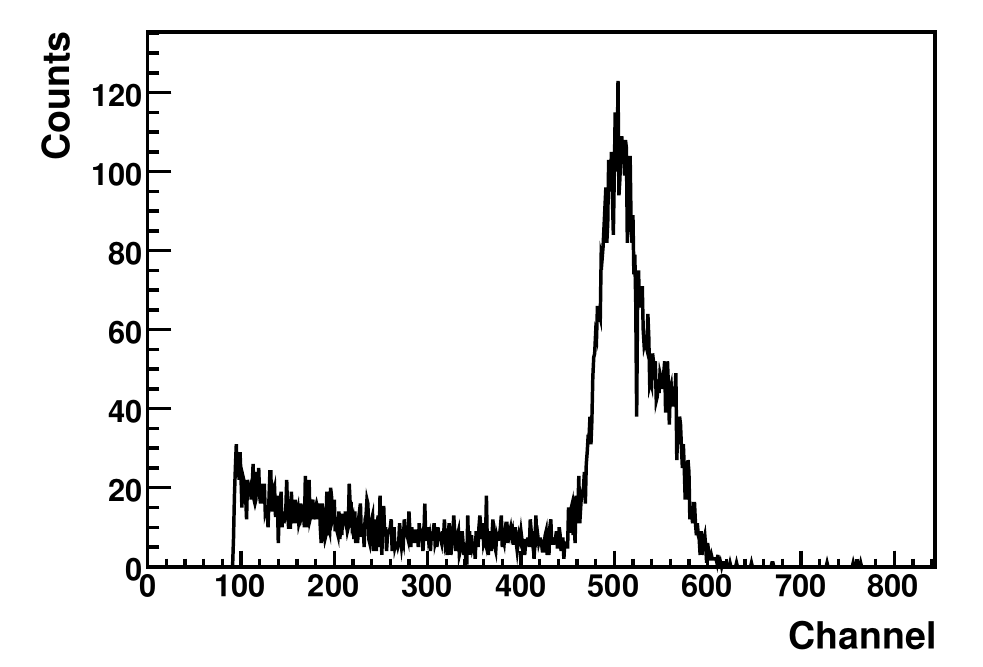}
\caption{Pulse height spectrum of the $^{55}$Fe source for a typical device at 207~K. The secondary bump on the upper side of the 5.90~keV peak is the 6.49~keV peak.}
\label{fig:resolution}
\end{figure}

\begin{figure}
\includegraphics[angle=90,width=78mm]{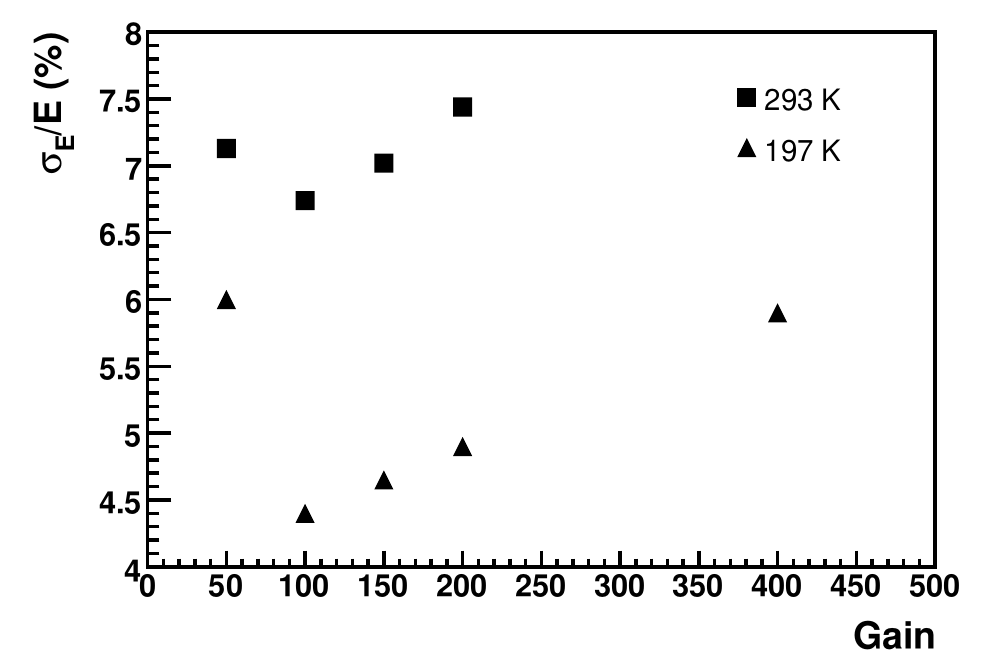}
\caption{Energy resolution of the 5.90~keV X-rays from the $^{55}$Fe source at 293~K and 197~K for a typical device. The best resolution is seen at gain of $\sim$100 for both temperatures.}
\label{fig:res_vs_gain}
\end{figure}

For accurate characterization of the LAAPDs gain and QE it is important to achieve good energy resolution in the $^{55}$Fe and XSLS spectra. Figure~\ref{fig:resolution} shows an example pulse height spectrum for the X-rays from the $^{55}$Fe source for a single LAAPD. The source produces X-rays of 5.90~keV (89\%) and of 6.49~keV (11\%).  The two peaks are not fully resolved even at low temperatures where the device noise is low. The energy resolution is determined by fitting the main peak of this spectrum to a Gaussian distribution. Figure~\ref{fig:res_vs_gain} shows the resulting energy resolution as a function of gain both close to room temperature and at 197~K.  It appears that the best resolution is obtained at a gain of $\sim$100 irrespective of the temperature.

The variance in the LAAPD output signal expressed in electrons, $\sigma_{N}^2$, is given by

\begin{equation}
\sigma_{N}^2 = G^2\sigma_{n}^2 + n_{\rm eh}\sigma_G^2 + n_{\rm eh}^2\sigma_{\rm NU}^2 + \sigma_{\rm noise}^2.
\label{eqn:sigmas}
\end{equation}

The first two terms describe the statistical variance in the number of primary photoelectrons and the gain respectively. Here $G$ is the gain, $\sigma_{n}^2$ is the variance in the number of primary photoelectrons, $n_{\rm eh}$ is the number of primary photoelectrons and $\sigma_G^2$ is the statistical variance in the single electron gain. In addition, $\sigma_{\rm NU}$ represents the gain non-uniformity in the diode volume and $\sigma_{\rm noise}$ is the dark noise of the diode-preamplifier system.

The variance in the primary electrons is determined by the Fano factor, $f$, with a value of about 0.1 in silicon~\cite{knoll00}

\begin{equation}
\sigma_n^2 = n_{\rm eh}f.
\label{eqn:primary}
\end{equation}

Following~\cite{moszynski02} we describe the statistical variance in the gain by introducing an excess noise factor, $F$

\begin{equation}
F=1+\frac{\sigma_G^2}{G^2},
\end{equation}

with $F=2.0$ at a gain of 100~\cite{moszynski00,ludhova05}. Thus Eqn.~\ref{eqn:sigmas} becomes

\begin{equation}
\sigma_{N}^2 = G^2[n_{\rm eh}f + n_{\rm eh}(F-1)] + n_{\rm eh}^2\sigma_{\rm NU}^2 + \sigma_{\rm noise}^2.
\end{equation}

This can be rewritten in terms of the energy released in the device, $E=n_{\rm eh}\epsilon$, where $\epsilon$ is average energy required to produce one e--h pair (3.66~eV for silicon). This leads to the equation for energy resolution, $\frac{\sigma_{E}}{E}$

\begin{equation}
\left(\frac{\sigma_{E}}{E}\right)^2 = \frac{\epsilon}{E}(f+F-1) + \frac{1}{G^2}\sigma_{\rm NU}^2 + \frac{\epsilon^2}{E^2G^2}\sigma_{\rm noise}^2.
\label{eqn:res}
\end{equation}

\begin{figure}
\includegraphics[angle=90,width=78mm]{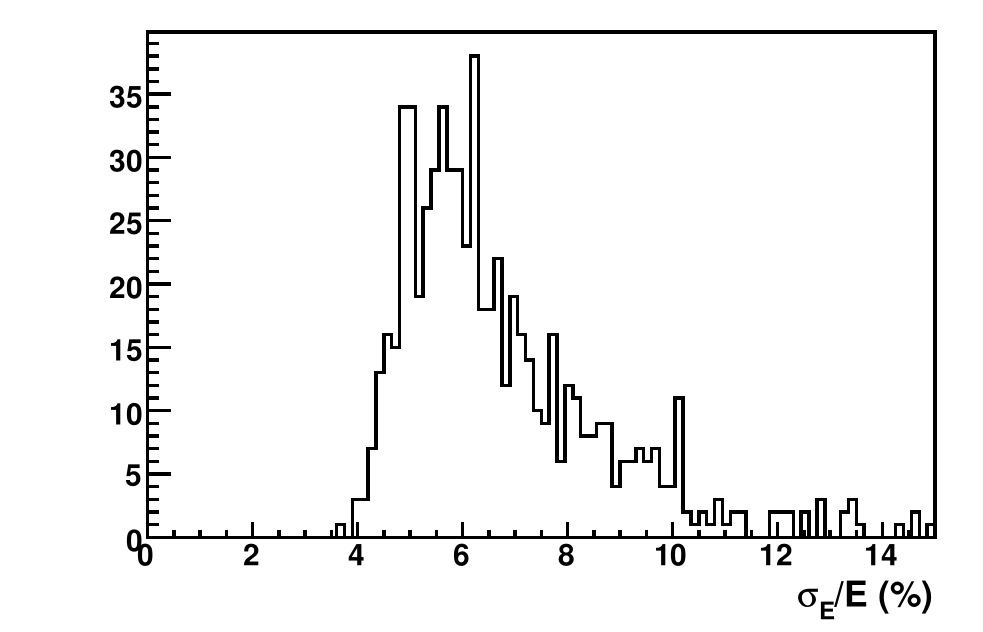}
\caption{Energy resolution of the 5.90~keV X-ray line for 667 EXO-200 production LAAPDs.}
\label{fig:apd_res}
\end{figure}

For 5.90~keV $^{55}$Mn X-rays, we calculate $\frac{\epsilon}{E}(f+F-1)\simeq(2.6\%)^2$. Figure~\ref{fig:apd_res} shows the observed energy resolution at 169~K for the production LAAPDs at a gain of $\sim$100. The dark noise term, $\sigma_{\rm noise}^2$, is known to be negligible at 169~K, so we interpret the peak energy resolution of $\sim$5.5\% as due to the non-uniformity term, $\sigma_{\rm NU}^2$, as was already noticed for the same devices in~\cite{moszynski00}.

Events from the XSLS consist of 150--300 VUV photons detected by each LAAPD, depending on the device location on the mounting disk. Since the photons strike the device at random locations gain non-uniformities are averaged out. However, the new terms: $\sigma_{\rm ph}$ accounting for the Poisson distribution in the number of detected photons, $\sigma_{\rm scint}$ accounting for the photon statistics in the scintillation process and $\sigma_{\rm source}$ deriving from the fact that the source is not point-like and hence different events are seen under slightly different solid angles, contribute to the variance in the primary electrons. Indeed, $^{148}$Gd was chosen as the $\alpha$ source because of its low energy (3.3~MeV), producing relatively localized scintillation tracks.

\begin{figure}
\includegraphics[angle=90,width=78mm]{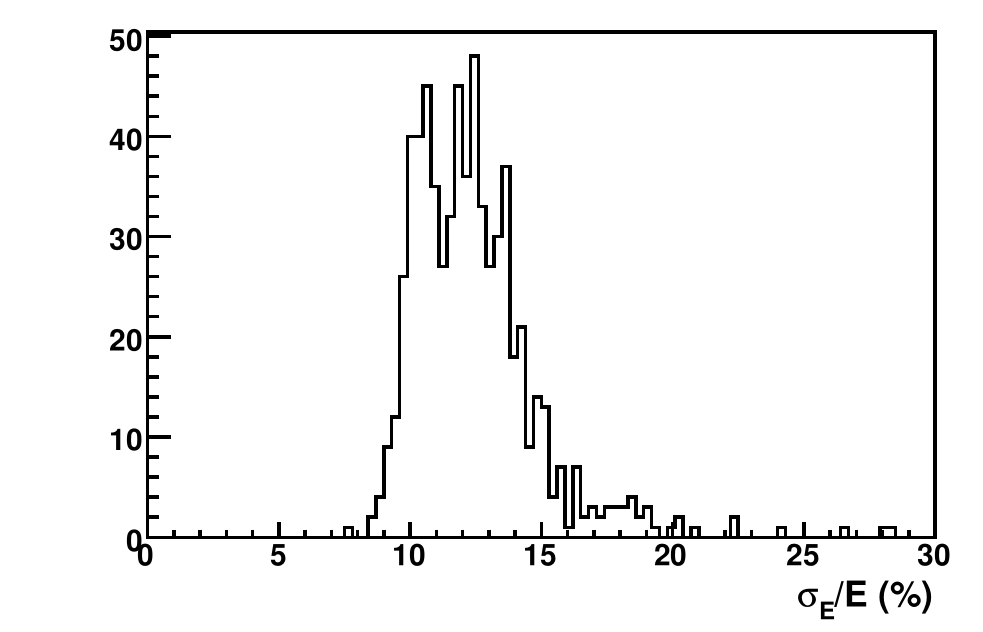}
\caption{Energy resolution of the XSLS events for 667 EXO-200 production LAAPDs.}
\label{fig:xsls_res}
\end{figure}

Figure~\ref{fig:xsls_res} shows the observed XSLS energy resolution at 169~K at a gain of $\sim$100. In most cases the observed energy resolution is in the range 9--15\%. LAAPDs tested in the central position of the mounting disk receive the most photons ($\sim$300 per event) and have the best XSLS resolution (9--10\%); LAAPDs tested in the outside positions receive fewer photons ($\sim$150 per event) and consequently have worse resolution (13--15\%). There are some devices with XSLS resolution above 15\%, but the majority of these are either particularly noisy or have unusually low QE. In fact, although XSLS resolution is not used as a selection criterion, only two devices eventually selected as suitable for installation in EXO-200 have XSLS resolution above 15.5\%, and none above 16.5\%.

\section{Gain}

\begin{figure}
\includegraphics[angle=90,width=78mm]{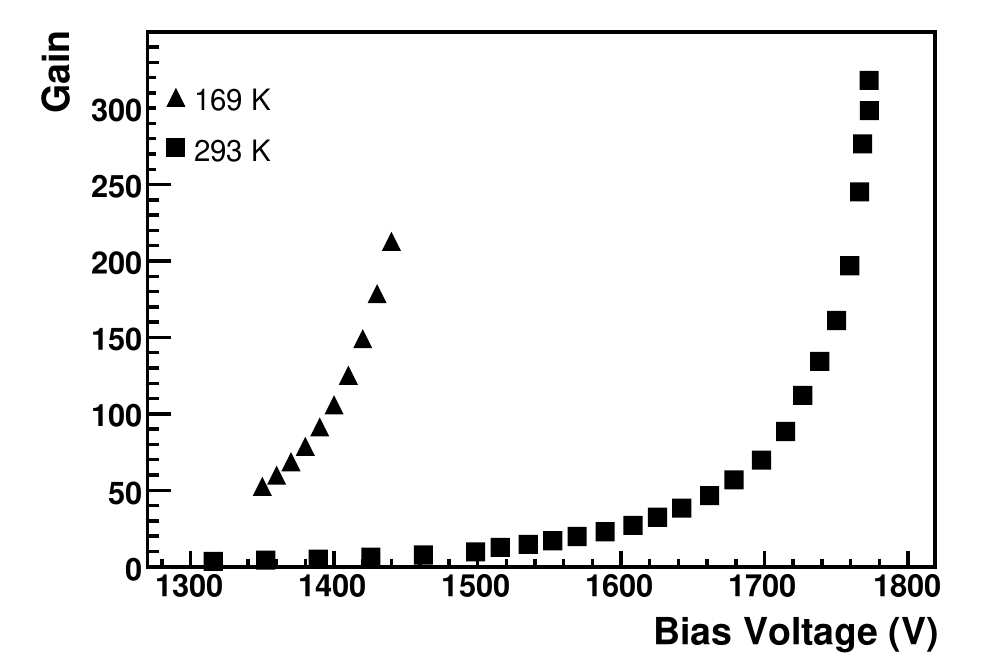}
\caption{Gain versus bias voltage near room temperature (data supplied by the manufacturer) and at 169~K (EXO data) for a typical device .}
\label{fig:gain_vs_voltage}
\end{figure}

\begin{figure}
\includegraphics[angle=90,width=78mm]{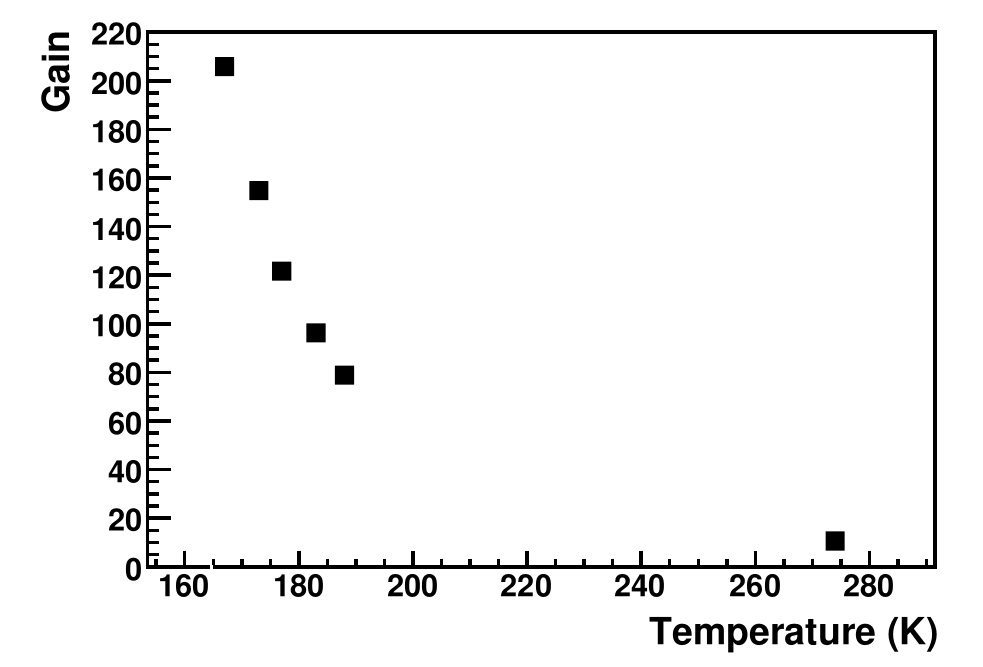}
\caption{Gain versus temperature at a fixed bias voltage (1439~V) for a typical device.}
\label{fig:gain_vs_temp}
\end{figure}

Figure~\ref{fig:gain_vs_voltage} and~\ref{fig:gain_vs_temp} show the gain of an LAAPD as a function of bias voltage and temperature. These data demonstrate the necessary requirements to maintain stable gain. At a gain of 100, corresponding to $\sim$1400~V bias, and at 169~K, the slope in the gain is approximately 1.5\%~V$^{-1}$. The gain also increases rapidly with decreasing device temperature, approximately 5\%~K$^{-1}$ near the operating temperature of 170~K. Hence to achieve a gain stability below 1\%, the variation in voltage (temperature) must be below 1~V (0.2~K).

\begin{figure}
\includegraphics[angle=90,width=78mm]{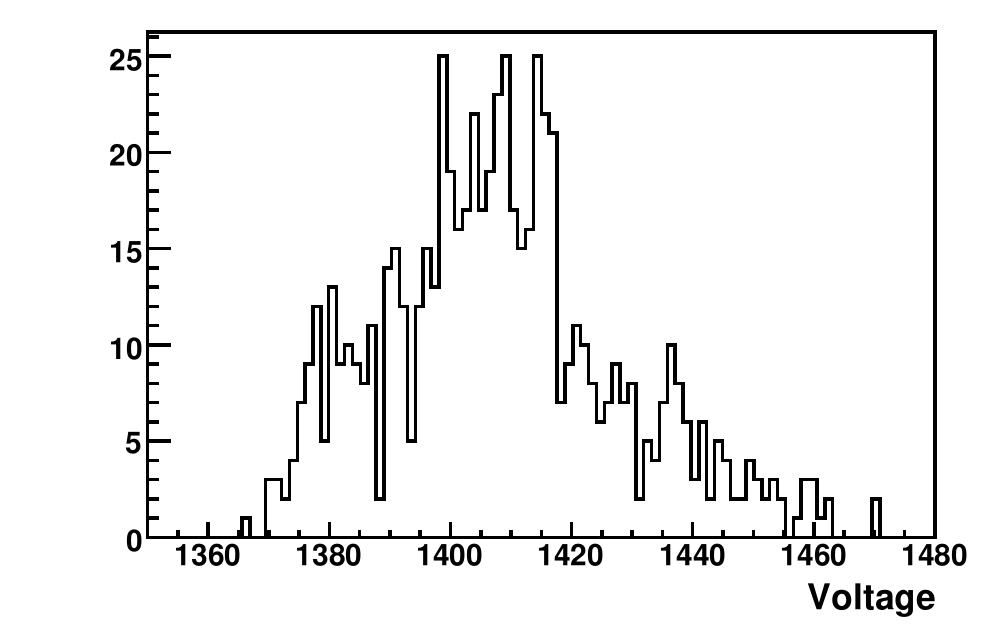}
\caption{Bias voltage required to operate at a gain of 100 for 667 EXO-200 production LAAPDs.}
\label{fig:apd_gain}
\end{figure}

Gain measurements are made for each production LAAPD at a series of bias voltages, typically over the approximate gain range 50-200. We construct a gain curve from these measurements, and determine the required bias voltage to operate at a gain of 100 for each LAAPD. Figure~\ref{fig:apd_gain} shows the distribution of voltages required to produce the same gain of 100 for all EXO-200 production LAAPDs. Gain uniformity in EXO-200 is obtained through a combination of device matching (in groups of $\sim$40 units with the same bias voltage) and bias voltage tuning (for each such group).

\section{VUV response}

We determine the relative QE for xenon scintillation light by measuring the response to XSLS events of each production LAAPD with respect to a reference LAAPD. Corrections are applied for the previously measured device gain and the location dependent XSLS brightness. The brightness at each of the 16 LAAPD locations is calibrated by comparing the response to the XSLS signal for a set of LAAPDs in multiple positions.

\begin{figure}
\includegraphics[angle=90,width=78mm]{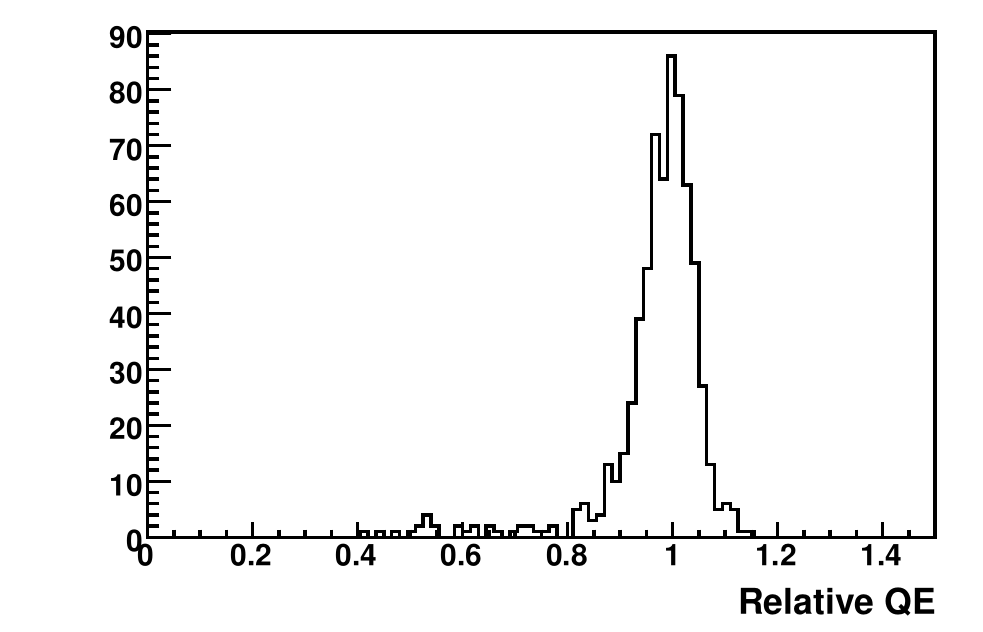}
\caption{Relative QE for 667 EXO-200 production LAAPDs.}
\label{fig:apd_qe}
\end{figure}

A histogram of the relative QE of the production LAAPDs is shown in Figure~\ref{fig:apd_qe}. We can see that the majority of the devices ($\sim$96\%) have relative QE between 0.8 and 1.2. The devices with relative QE below 0.8 are not used in EXO-200.

Surface cleanliness is critical for maintaining good VUV response, and we have observed the effects of surface contamination on two sets of 15 devices. Significant deterioration in VUV QE for these 30 devices was noticed following accidental high voltage discharges in the test chamber. The devices were otherwise unaffected, and in most cases their response to VUV light partially recovered after ultrasonic cleaning in methanol and deionized water followed by a vacuum bake at 160$^\circ$~C for 24 hours. This suggests that surface contamination was indeed the cause of the poor VUV response, although the nature of the contamination and the mechanism by which it was released during the electrical breakdown episodes remains unclear.

\section{Stability}

\begin{figure}
\includegraphics[angle=90,width=78mm]{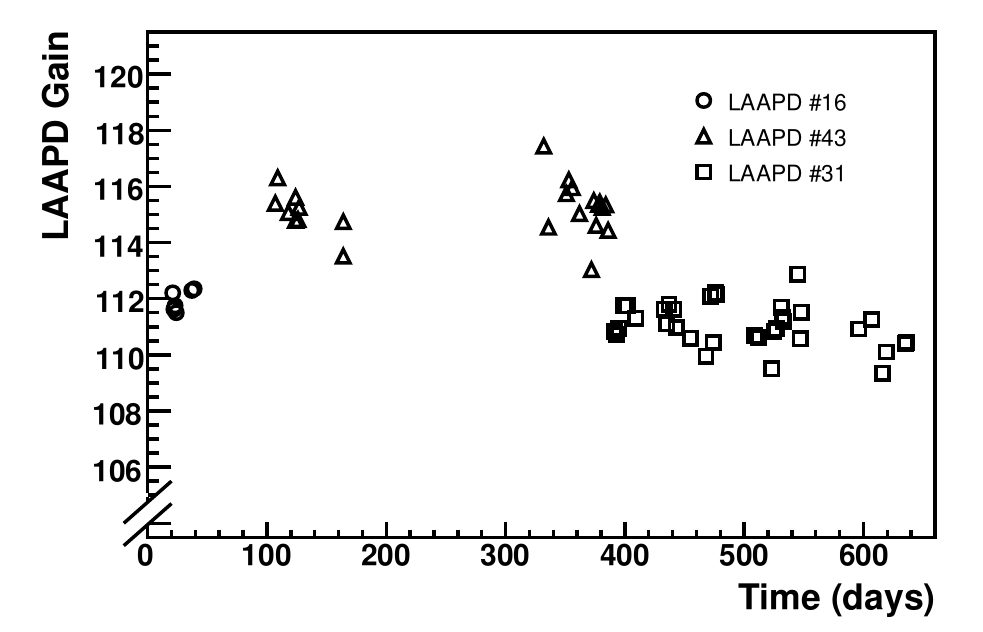}
\caption{Variation of reference LAAPD gain at a bias of 1420~V and at 169~K. Note the zero suppression. The horizontal axis shows time since testing began in November 2006. Three different reference LAAPDs were used over the course of two years of testing, each with slightly different gains---explaining the jumps at days 100 and 390---but for each of the three no systematic gain fluctuations are observed.}
\label{fig:gainvtime}
\end{figure}

The stability of the multiple-LAAPD test setup is monitored by keeping the same device in the central position over many testing cycles. These central LAAPDs (a total of three have been used over the course of two years of testing), as well as several others that were characterized a second time several months after the initial characterization, also allow us to measure the stability of the devices themselves over time. Figure~\ref{fig:gainvtime} displays the measured gain of the central LAAPDs over time. No systematic changes are evident and the gain of each device is constant to within 2\%. 

\section{Conclusions}

We have performed a systematic set of measurements on 851 bare, low activity LAAPDs at room temperature and at about 170~K. We find that 565 devices comply with the EXO-200 requirements for use as VUV scintillation light detectors in the experiment 
(468 devices are needed). Most of the devices rejected have either high dark noise ($\sim$250 devices) or low relative QE ($\sim$30 devices). 

\section{Acknowledgments}

We would like to thank the staff of Advanced Photonix for their cooperation in what turned out to be a rather exotic enterprise.  We are also grateful to M. Szawlowski for his advice and guidance and to International Rectifiers for providing---on very short notice---epitaxial wafers at no cost. The Teflon used for the XSLS was provided free of charge by the DuPont corporation. EXO is supported by DoE and NSF in the United States, NSERC in Canada, FNS in Switzerland and the Ministry for Science and Education of the Russian Federation in Russia.



\end{document}